\documentclass[twocolumn,prl,aps,superscriptaddress,floatfix]{revtex4-1}
\usepackage{graphicx}
\usepackage{amsmath}
\usepackage{natbib}
\usepackage{gensymb}

\begin{document}

\title{Electrical switching of a moir\'{e} ferroelectric superconductor}

\date{{\small \today}}

\author{Dahlia R. Klein}
\email{dahlia.klein@weizmann.ac.il}
\affiliation{Department of Physics, Massachusetts Institute of Technology, Cambridge, Massachusetts 02139, USA}
\affiliation{Department of Condensed Matter Physics, Weizmann Institute of Science, Rehovot 7610001, Israel}
\author{Li-Qiao Xia}
\author{David MacNeill}
\affiliation{Department of Physics, Massachusetts Institute of Technology, Cambridge, Massachusetts 02139, USA}
\author{Kenji Watanabe}
\author{Takashi Taniguchi}
\affiliation{National Institute for Materials Science, Namiki 1-1, Tsukuba 305‐0044, Japan}
\author{Pablo Jarillo-Herrero}
\email{pjarillo@mit.edu}
\affiliation{Department of Physics, Massachusetts Institute of Technology, Cambridge, Massachusetts 02139, USA}

\maketitle

\textbf{Electrical control of superconductivity is critical for nanoscale superconducting circuits including cryogenic memory elements \cite{Baek14,Gingrich16,Sardashti20,Alam21}, superconducting field-effect transistors (FETs) \cite{Doh05,DeSimoni18,Fatemi18}, and gate-tunable qubits \cite{Larsen15,deLange15,Wang19}. Superconducting FETs operate through continuous tuning of carrier density, but there has not yet been a bistable superconducting FET, which could serve as a new type of cryogenic memory element. Recently, unusual ferroelectricity in Bernal-stacked bilayer graphene aligned to its insulating hexagonal boron nitride (BN) gate dielectrics was discovered \cite{Zheng20}. Here, we report the observation of ferroelectricity in magic-angle twisted bilayer graphene (MATBG) with aligned BN layers. This ferroelectric behavior coexists alongside the strongly correlated electron system of MATBG without disrupting its correlated insulator or superconducting states. This all-van der Waals platform enables configurable switching between different electronic states of this rich system. To illustrate this new approach, we demonstrate reproducible bistable switching between the superconducting, metallic, and correlated insulator states of MATBG using gate voltage or electric displacement field. These experiments unlock the potential to broadly incorporate this new moir\'{e} ferroelectric superconductor into highly tunable superconducting electronics.}

Twisting two layers of graphene forms a moir\'{e} superlattice with alternating regions of AA and AB/BA stacking. Near the magic angle of 1.1$\degree$, interlayer hybridization at the AA regions leads to renormalized flat energy bands \cite{Suarez10,Bistritzer11,Lopes12,Li10,Luican11}, thereby suppressing the electronic kinetic energy and enabling strong Coulomb interactions to dominate. Experimental realizations of MATBG have led to a number of surprising results including the discovery of superconductivity \cite{Cao18b}, correlated insulating states \cite{Cao18a}, orbital magnetism and the quantum anomalous Hall effect \cite{Sharpe19,Serlin20}, and strange metal behavior \cite{Cao20}. Beyond MATBG, a new field of twistronics has begun to explore the effects of moir\'{e} superlattices in other van der Waals crystals. Researchers have investigated moir\'{e} patterns in twisted bilayers of BN at small twist angles \cite{Yasuda21,Vizner21}, resulting in ferroelectricity (FE) due to a shift in the ionic dipole from the boron and nitrogen ions. Surprisingly, ferroelectricity has also been observed in a double moir\'{e} system from the alignment of the BN layers sandwiching Bernal-stacked AB bilayer graphene \cite{Zheng20}. In this work, 
we present the discovery of ferroelectric behavior in MATBG, in which the top and bottom BN crystallographic edges are closely aligned to one another modulo 30$\degree$.

\begin{figure*}[!tbp]
\includegraphics[width=11 cm]{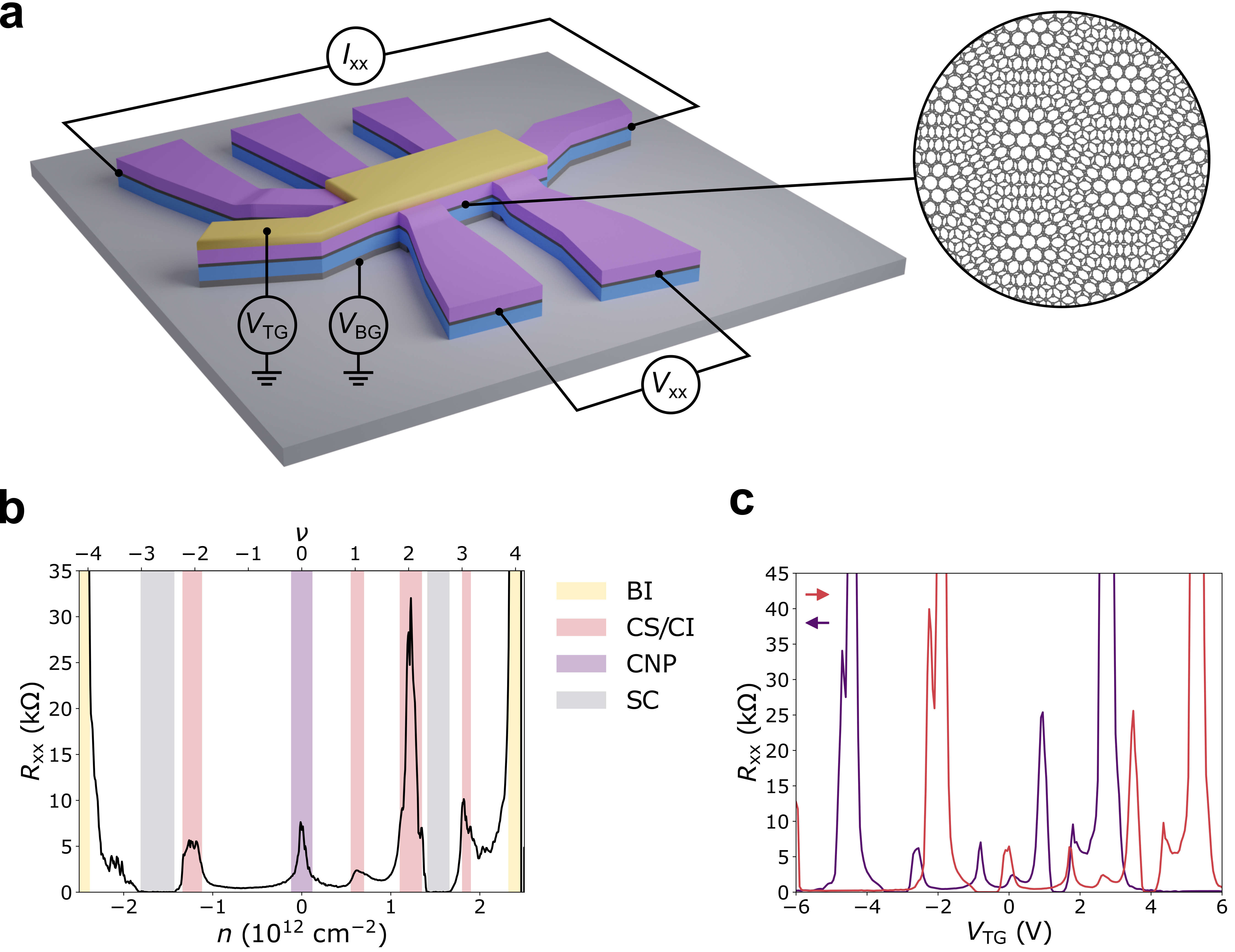}
    \caption{\textbf{Device characterization.} \textbf{(a)} Schematic of the dual-gated MATBG device in a Hall bar geometry using a metal top gate and few-layer graphite bottom gate. \textbf{(b)} Longitudinal resistance $R_\mathrm{xx}$ as a function of carrier density $n$. The regions of band insulator (BI), correlated semimetal/insulator (CS/CI), charge neutrality point (CNP), and superconductivity (SC) are highlighted. \textbf{(c)} Hysteresis of $R_\mathrm{xx}$ vs. applied top gate voltage $V_\mathrm{TG}$ depending on gate sweep direction.}
    \label{fig1}
\end{figure*}

The stack was fabricated using standard cut-and-stack dry-transfer van der Waals assembly (see Methods). The final device, shown schematically in Fig. 1a, is etched into a Hall bar geometry with a metal (Cr/Au) top gate and local few-layer graphite bottom gate. During van der Waals assembly, the long, sharp crystallographic edges of the top and bottom BN flakes were intentionally aligned to one another, resulting in the two BN layers approximately aligned modulo 30$\degree$ (see Fig. S1).

Low-temperature transport measurements demonstrate the expected features of a high-quality twisted bilayer graphene device close to the magic angle $\theta\approx 1.1\degree$. In Fig. 1b, the four-probe longitudinal resistance $R_\mathrm{xx}$ is plotted as a function of carrier density using the local graphite bottom gate ($V_\mathrm{BG}$) at a base mixing chamber temperature of 40 mK. The filling factor $\nu$ refers to the number of electrons or holes per MATBG moir\'{e} superlattice unit cell; $\nu$ = $\pm$4 refer to full filling of the moir\'{e} bands with four electrons or holes per unit cell, respectively. The transport data show prominent resistive features corresponding to the charge neutrality point (CNP, $\nu$ = 0), band insulating peaks (BI, $\nu$ = $\pm$4), and correlated semimetal/insulating peaks (CS/CI, $\nu$ = 1, $\pm$2, 3). Moreover, robust superconducting (SC) regions appear near half-filling of the moir\'{e} unit cell at $\nu$ = $\pm (2+\delta)$. A twist angle of 1.03$\degree$ can be extracted from the CNP and $\nu$ = 2 peak (see Methods).

Strikingly, when the top gate is instead used to tune the carrier density in the device, there is a large hysteretic shift in the transport features of MATBG depending on the sweep direction of $V_\mathrm{TG}$ (Fig. 1c). When the gate voltage is initially swept, there is a large region over which the gate appears to not work, followed by a region of expected behavior where the MATBG carrier density is changing. This observation is consistent with the proposed picture of ``layer-specific anomalous screening" (LSAS) described in Ref. \cite{Zheng20}.

To further investigate this emergent ferroelectric behavior in our MATBG device, we performed dual-gate maps of $R_\mathrm{xx}$ as a function of both gate voltages. We show the effects of sweeping the top gate $V_\mathrm{TG}$ as the slow axis either from negative to positive voltage (Fig. 2a) or from positive to negative voltage (Fig. 2b). When $V_\mathrm{TG}$ is initially changed, the transport features gradually evolve, suggesting that the gate is not fully introducing carriers into the MATBG layers as would be expected in a standard field-effect transistor. Upon further modulation of $V_\mathrm{TG}$, the device abruptly enters the familiar regime where the resistive peaks, which occur at constant carrier density in MATBG, follow straight-line trajectories in the dual-gate plane. This change in behavior can be visualized more easily using converted axes in Fig. 2c (converted from Fig. 2a) and 2d (converted from Fig. 2b). The transformation is defined as: $n_\mathrm{ext} = (\varepsilon_\mathrm{BN}/e)(V_\mathrm{TG}/d_\mathrm{TG}+V_\mathrm{BG}/d_\mathrm{BG})$, $D_\mathrm{ext}/\varepsilon_0 = (\varepsilon_\mathrm{BN}/2)(V_\mathrm{TG}/d_\mathrm{TG}-V_\mathrm{BG}/d_\mathrm{BG})$. In the regions where the top gate is adding carriers to the MATBG device by the expected amount, the transport features follow vertical-line trajectories, in agreement with previous observations that the strongly correlated electron system in MATBG is independent of displacement field $D_\mathrm{ext}$ \cite{Sharpe19}.

Based on our dual-gate observations in Fig. 2, we conclude that the emergent ferroelectricity in our device persists alongside the strongly correlated behavior of MATBG. To confirm this coexistence, we study the superconductivity in our device in one of its two bistable configurations. In Fig. 3a, we plot the four-probe $R_\mathrm{xx}$ as a function of bottom gate and temperature. Two superconducting domes appear at $\nu$ = $\pm(2+\delta)$, in agreement with prior MATBG phenomenology. Using the definition of critical temperature as 50\% of the normal state resistance (see Supplemental Information), we extract maximal $T_\mathrm{c}$ of 2.15 K ($\nu=-2.62$) and 0.83 K ($\nu=2.32$) for the hole and electron domes, respectively.

\begin{figure*}
\includegraphics[width=13 cm]{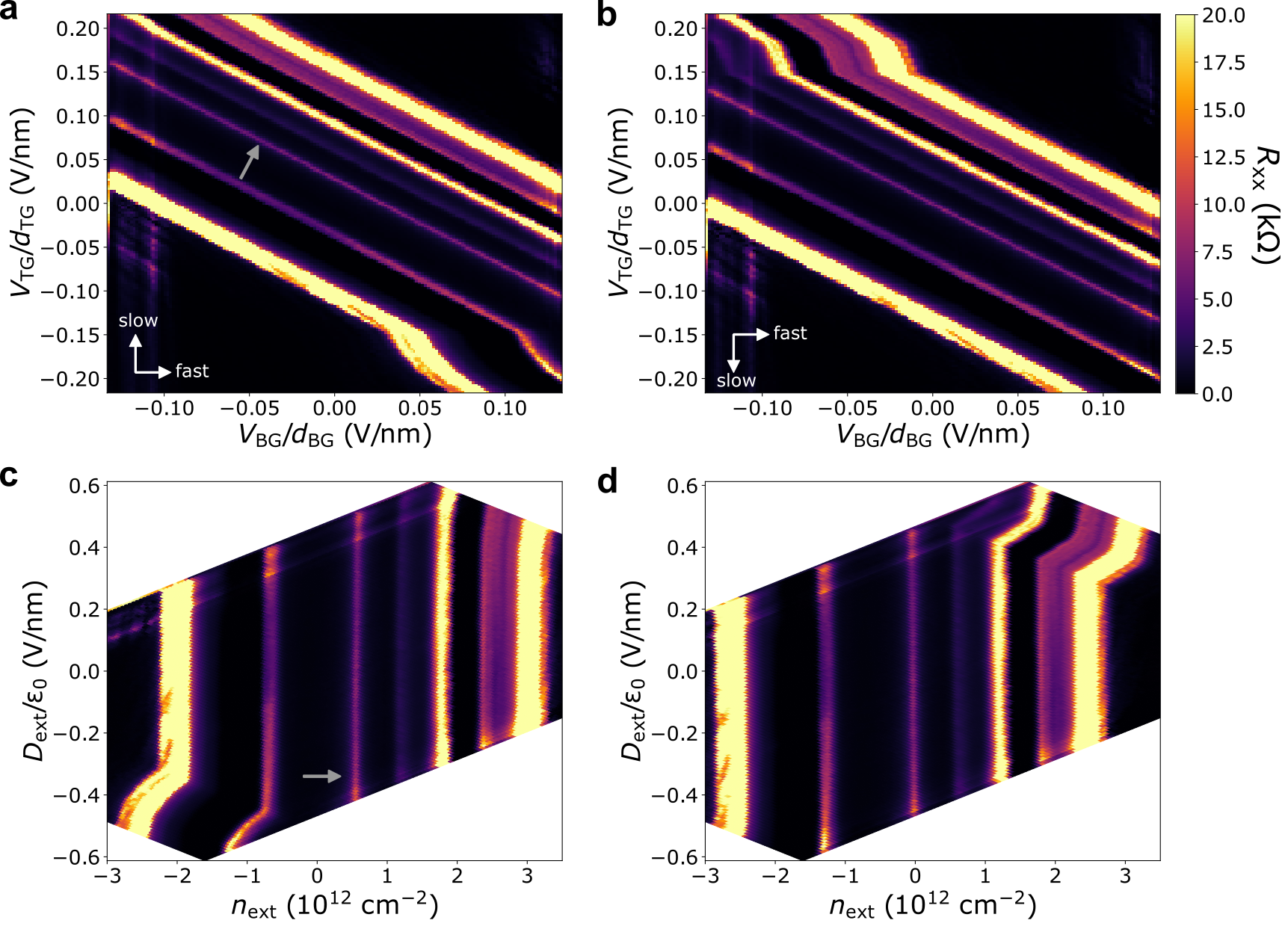}
    \caption{\textbf{Dual-gate maps of longitudinal resistance.} \textbf{(a, b)} Four-probe longitudinal resistance $R_\mathrm{xx}$ vs. top gate $V_\mathrm{TG}$ and bottom gate $V_\mathrm{BG}$. The fast scan axis is $V_\mathrm{BG}$, swept up from negative to positive. The slow scan axis is $V_\mathrm{TG}$, swept up in \textbf{(a)} and down in $V_\mathrm{TG}$. Axes are normalized to the BN dielectric thicknesses of the two gates. \textbf{(c, d)} Converted maps of \textbf{(a, b)} in units of $n_\mathrm{ext}$ and $D_\mathrm{ext}/\varepsilon_0$ (see main text). The arrows in \textbf{(a)} and \textbf{(c)} indicate the CNP.}
    \label{fig2}
\end{figure*}

The superconductivity remains prominent over large regions on both the electron and hole sides at $\nu$ = $\pm(2+\delta)$ (Fig. 3b). Using the bottom gate, we can park the system in either region at fixed carrier density. In Fig. 3c, we show the differential resistance $dV_\mathrm{xx}/dI$ as a function of both DC current $I_\mathrm{DC}$ and applied perpendicular magnetic field $B_\perp$ on the hole side at $\nu=-2.48$. We find that the superconducting critical current is maximized at $B_\perp$ = 0 and falls to zero with increasing $B_\perp$ approaching $\pm 100$ mT. We find similar results on the electron side at $\nu=2.57$ (Fig. 3d), with the critical current vanishing near $B_\perp$ = $\pm 40$ mT. These data are hallmarks of robust superconductivity persisting in MATBG and, importantly, are not influenced by the coexisting FE behavior. The underlying mechanism of the ferroelectricity now observed in both Bernal bilayer graphene and MATBG sandwiched by aligned BN dielectrics is still not understood. These results in MATBG provide valuable evidence in developing an understanding of where the ferroelectricity lies, given its clear lack of interplay with the strongly interacting electrons in the MATBG moir\'{e} superlattice.

The FE hysteresis persists upon cycling between the two hysteretic sweep directions. In Fig. 4a, $R_\mathrm{xx}$ is shown when the top gate $V_\mathrm{TG}$ is swept back and forth from $\pm$6 V over six traces. A clear bistability emerges: the up and down traces occur at the same positions depending only on sweep direction. Furthermore, the history of the applied top gate also influences the bottom gate sweeps. We can prepare the system in either state by setting $V_\mathrm{BG}$ = 0, applying $V_\mathrm{TG}$ = $\pm$6 V for several seconds, and then setting $V_\mathrm{TG}$ back to 0. Next, we set $V_\mathrm{BG}$ = $\mp$4 V and sweep the bottom gate. This sign is chosen to maintain the same sign of $D_\mathrm{ext}$ at the start of each scan. As shown in Fig. 4b, the system is again bistable when this sequence is repeated over six traces.

\begin{figure*}[!tbp]
\includegraphics[width=14 cm]{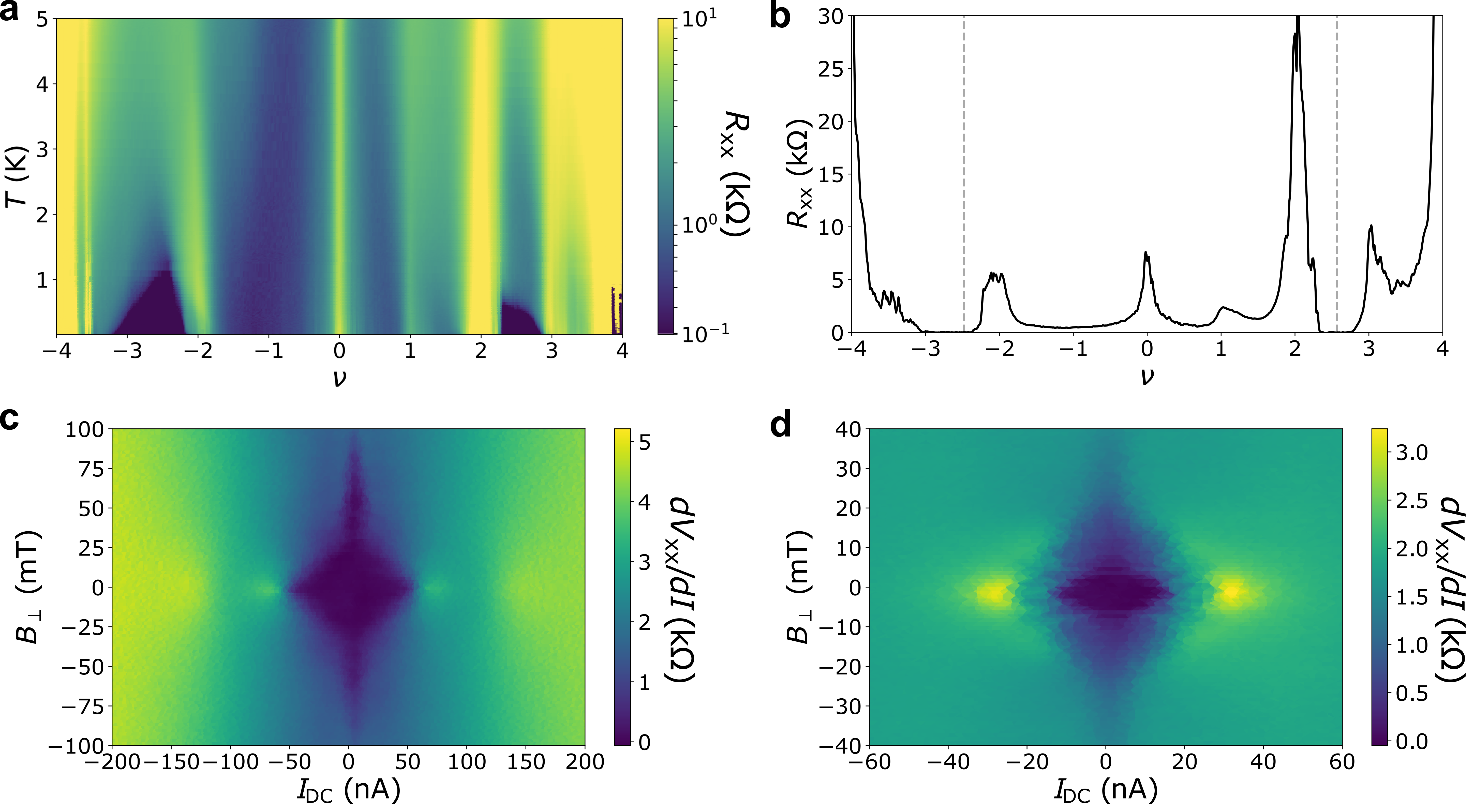}
    \caption{\textbf{Characterization of robust superconductivity.} \textbf{(a)} Temperature dependence of the four-probe longitudinal resistance $R_\mathrm{xx}$ vs. $V_\mathrm{BG}$. \textbf{(b)} $R_\mathrm{xx}$ vs. $V_\mathrm{BG}$ at a mixing chamber $T$ = 65 mK. The dashed lines show the positions of the maps taken in \textbf{(c,d)} at $\nu=-2.48$ and $\nu=2.57$, respectively. \textbf{(c,d)} Differential resistance $dV_\mathrm{xx}/dI$ maps vs. DC current $I_\mathrm{DC}$ and applied perpendicular magnetic field $B_\perp$ on the hole \textbf{(c)} and electron \textbf{(d)} sides.}
    \label{fig3}
\end{figure*}

Now that we have established the coexistence of a FE bistability and strongly correlated MATBG behavior in a single device, we can make use of this property to reversibly switch between different electronic states. In Fig. 4c, we initially prepare the system with a pulse of $V_\mathrm{TG}$ = $-$6 V before setting $V_\mathrm{TG}$ = 0. We then fix the applied bottom gate at $-$1.4 V, placing the system in the hole-side ($\nu=-2-\delta$) superconducting phase. Remarkably, we demonstrate switching of $R_\mathrm{xx}$ from this superconducting state to a stable metallic state after applying a pulse of $V_\mathrm{TG}$ = 6 V and setting the top gate back to 0 (Fig. 4c). After pausing in the metallic state, we then apply a second pulse of $V_\mathrm{TG}$ = $-$6 V to return the device to the superconducting state. In order to emphasize the stability of this procedure, we also perform a pulsed sequence of alternating $V_\mathrm{TG}$ = $\pm$6 V pulses to repeatably switch the system between the bistable superconducting and metallic states at a fixed bottom gate voltage (Fig. 4d).

Moreover, we also demonstrate the capability to switch between different electronic phases of MATBG at fixed applied carrier density using only applied displacement field. Employing the definitions of $n_\mathrm{ext}$ and $D_\mathrm{ext}/\varepsilon_0$ stated earlier, we fix the system at $n_\mathrm{ext}$ = $-8.37\cdot10^{11}$ $\mathrm{cm}^{-2}$. Next, using alternating pulses of $D_\mathrm{ext}/\varepsilon_0$ = 0.189 V/nm and -0.379 V/nm, we show the capacity to switch between superconducting and resistive correlated insulator states (Fig. 4e). This repeatability reflects the robust FE behavior existing in this system and illustrates its utility as a novel van der Waals platform for superconducting switches.

\begin{figure}[!tbp]
\includegraphics[width=9 cm]{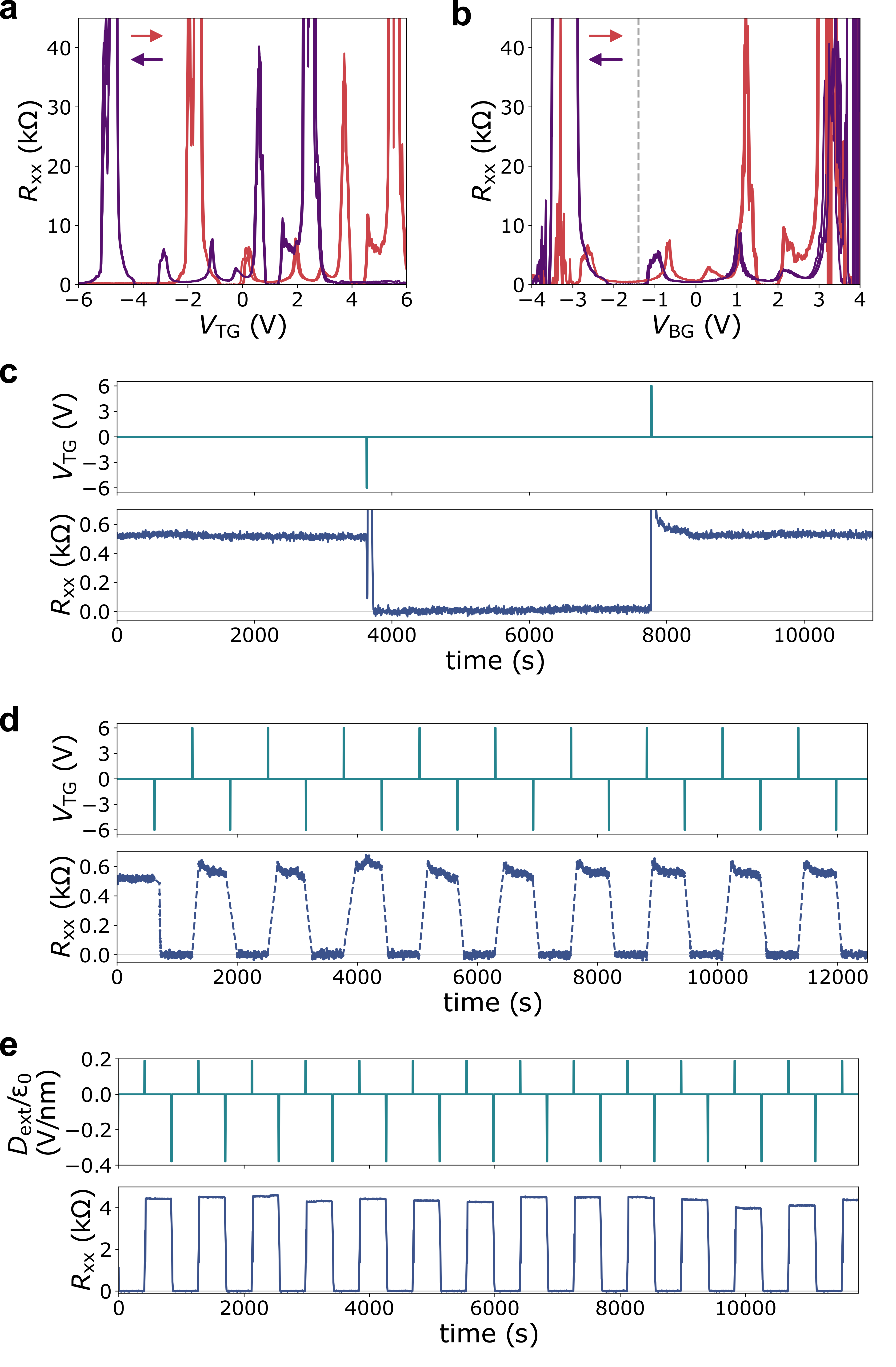}
    \caption{\textbf{Ferroelectric switching of MATBG states and superconductivity.} \textbf{(a)} $R_\mathrm{xx}$ vs. $V_\mathrm{TG}$ with sweep directions up (pink) and down (purple) for $V_\mathrm{BG}$ = 0. \textbf{(b)} $R_\mathrm{xx}$ vs. $V_\mathrm{BG}$ with sweep directions up (pink) and down (purple) for $V_\mathrm{TG}$ = 0. In between traces, $V_\mathrm{BG}$ was set to 0, $V_\mathrm{TG}$ was increased to $\pm$6 V and then back to 0 before starting a new trace at $V_\mathrm{BG}$ = $\mp$4 V. The dashed line denotes $V_\mathrm{BG}$ = $-$1.4 V. \textbf{(c)} $R_\mathrm{xx}$ vs. time with fixed $V_\mathrm{BG}$ = $-$1.4 V. A top gate voltage $V_\mathrm{TG}$ = $\pm$6 V was applied to switch between metallic and superconducting states, respectively. \textbf{(d)} Illustration of reversible switching of $R_\mathrm{xx}$ at fixed $V_\mathrm{BG}$ = $-$1.2 V over many cycles. \textbf{(e)} Reversible switching using only applied displacement field $D_\mathrm{ext}$ at a fixed carrier density of $n_\mathrm{ext}$ = $-8.37\cdot10^{11}$ $\mathrm{cm}^{-2}$. $R_\mathrm{xx}$ shows switching between correlated insulating and superconducting phases after positive and negative applied $D_\mathrm{ext}$ pulses, respectively. All data were taken at a mixing chamber temperature of 65 mK.}
    \label{fig4}
\end{figure}

Past research on controlling superconductivity has primarily focused on FETs to continuously tune carrier density, thereby inducing superconductivity in oxide thin films \cite{Caviglia08,Ueno08} and ultrathin van der Waals crystals \cite{Ye12,Fatemi18,Cao18b}. However, these devices rely on a continuous change in an applied gate voltage rather than a discrete switch between two stable states. Furthermore, they typically require much larger carrier densities ($\approx 10^{12}-10^{13}$ cm$^{-2}$) than the low-density SC in MATBG ($<10^{12}$ cm$^{-2}$) owing to the large area of the MATBG moir\'{e} unit cell. Other types of experiments have successfully turned superconductivity on and off using light pulses \cite{Fausti11,Yang19} or applied magnetic fields \cite{Taniguchi03}, but these are limited in their applicability compared to electrical control. Previous work employing both ferroelectricity and superconductivity using thin film FE oxides show modulation of $T_\mathrm{c}$ with FE polarization, but do not report controllable switching behavior between superconducting and other electronic phases at fixed temperature \cite{Ahn99,Takahashi06}. The data presented in this letter thus mark the first example of a nonvolatile FE SC FET.

To summarize, our results introduce a highly tunable all-van der Waals platform to enact bistable ferroelectric switching between a range of electronic states in a strongly correlated electron system at low carrier density. Recent progress in moir\'{e} superconducting devices using patterned electrostatic gating of MATBG with BN dielectrics has achieved Josephson junctions \cite{Rodan21,deVries21}, including magnetic Josephson junctions \cite{Diez21}, superconducting diodes \cite{Diez21}, and superconducting quantum interference devices (SQUIDS) \cite{Portoles22}. These van der Waals heterostructures, constructed from a single material platform, bypass the issues of strain and disorder that often occur at the interfaces between different thin films. Combining the bistable ferroelectric switching demonstrated here with configurable Josephson junction geometries will enable an additional control knob over the electronic states, paving the way for a new generation of moir\'{e} graphene superconducting electronics.

\hfill \break
\newline
\noindent \textbf{Methods}

\textbf{Device fabrication.} The van der Waals heterostructure was assembled in two parts using standard dry transfer techniques. First, a poly(bisphenol A carbonate) stamp was used to pick up a hexagonal boron nitride (BN) flake and few-layer graphite strip. This bottom gate stack was placed on a 285 nm SiO$_2$/Si substrate and annealed to remove any residue. Next, a large monolayer graphene flake was cut into two pieces using a laser setup described in Ref. \cite{Park21}. A second stamp was used to pick up a top BN flake using heat. At room temperature, the stamp then picked up one monolayer, the stage was rotated by 1.1$\degree$, and then the stamp picked up the other half. Finally, the stack was released onto the bottom gate stack with alignment of the long, sharp edges of the two BN layers serving as the gate dielectrics. These edges are likely either zigzag or armchair terminations of the crystal lattice. Thus, the BN layers are crystallographically aligned modulo 30$\degree$.

A metallic top gate of Cr/Au (5/40 nm) was deposited and the device was etched into a Hall bar geometry using reactive ion etching. The MATBG layers were contacted using a combined etch and evaporation of Cr/Au (5/40 nm) metallic contacts.

\textbf{Transport measurements.} Low-temperature electrical transport measurements were carried out in a dilution refrigerator with a perpendicular superconducting magnet. The sample current and four-probe voltage were measured using SR-830 lock-in amplifiers with pre-amplifier gains of $10^6$ and 10, respectively. The lock-ins were synchronized at a frequency of 7-8 Hz and an AC excitation current of 1 nA was applied. For the temperature dependent measurements in Fig. 3a and supplemental data, an on-chip thermometer (Lakeshore RX-102A-BR) was employed.

\textbf{Twist angle determination.} The twist angle of the MATBG device can be extracted from the positions of the insulating $R\mathrm{xx}$ peaks as a function of carrier density using the bottom gate. We can obtain the carrier density $n$ = ($\varepsilon_\mathrm{BN}/e$)($V_\mathrm{BG}$/$d_\mathrm{BG}$), where we take $\varepsilon_\mathrm{BN}$ = 3.5$\varepsilon_0$. Full filling of the moir\'{e} superlattice occurs for $n_{\nu=\pm4} = \pm8\sin^{2}\theta/\sqrt{3}a^2 \approx
\pm8\theta^2/\sqrt{3}a^2$, where the lattice constant $a$ = 0.246 nm for graphene. The value $n_{\nu=4}$ was calculated using the positions of insulating peaks at $\nu$ = 0 and 2.

\hfill \break
\noindent \textbf{Acknowledgements}

The authors acknowledge Qiong Ma and Zhiren Zheng for helpful discussions. This work was supported by the Air Force Office of Scientific Research (AFOSR) 2DMAGIC MURI FA9550-19-1-0390 (D.R.K. and L.-Q.X.), the Army Research Office MURI W911NF2120147 (D.M.), as well as the Gordon and Betty Moore Foundation's EPiQS Initiative through grant GBMF9463 to P.J.-H. K.W. and T.T. acknowledge support from the Elemental Strategy Initiative conducted by the MEXT, Japan (JPMXP0112101001), JSPS KAKENHI (JP20H00354), and the CREST(JPMJCR15F3), JST. This work made use of the MIT MRSEC Shared Experimental Facilities, supported by the NSF (DMR-0819762), and of Harvard's Center for Nanoscale Systems, supported by the NSF (ECS-0335765).

\bibliography{main}
 
\end{document}

\title{Supplementary Information for \\
Electrical switching of a moir\'{e} ferroelectric superconductor}

\date{{\small \today}}

\author{Dahlia R. Klein}
\email{dahlia.klein@weizmann.ac.il}
\affiliation{Department of Physics, Massachusetts Institute of Technology, Cambridge, Massachusetts 02139, USA}
\affiliation{Department of Condensed Matter Physics, Weizmann Institute of Science, Rehovot 7610001, Israel}
\author{Li-Qiao Xia}
\author{David MacNeill}
\affiliation{Department of Physics, Massachusetts Institute of Technology, Cambridge, Massachusetts 02139, USA}
\author{Kenji Watanabe}
\author{Takashi Taniguchi}
\affiliation{National Institute for Materials Science, Namiki 1-1, Tsukuba 305‐0044, Japan}
\author{Pablo Jarillo-Herrero}
\email{pjarillo@mit.edu}
\affiliation{Department of Physics, Massachusetts Institute of Technology, Cambridge, Massachusetts 02139, USA}

\maketitle

\tableofcontents

\newpage

\begin{section}{Crystallographic alignment of BN flakes}

During van der Waals assembly, the long crystallographic edges of the top and bottom BN flakes were aligned to each other. These edges are typically zigzag or armchair terminations of the lattice. Therefore, the edges are known modulo 30$\degree$. In Fig. S1, an optical microscope image of the final heterostructure is shown after removal of polycarbonate residues with a chloroform soak. The extracted relative BN alignment angle is $1.4\pm0.1\degree$ (modulo 30$\degree$). In order to minimize relaxation of the moir\'{e} heterostructures, no heat annealing steps were performed post-transfer.

Our device displays the same hysteresis direction as the reported device (`Device H2') of Ref. \cite{Zheng20} consisting of AB-stacked bilayer graphene with its crystallographic axis aligned to the two BN axes. Second-harmonic generation (SHG) experiments reveal that their two BN crystals are at a 30$\degree$ angle to one another (modulo 60$\degree$), in contrast to the reversed hysteresis direction for a different device with 0$\degree$ alignment (modulo 60$\degree$). On this basis, we hypothesize that our MATBG device also has relative BN alignment of 30$\degree$.

\vspace{1cm}

\begin{figure*}[!htb]
\includegraphics[width=9 cm]{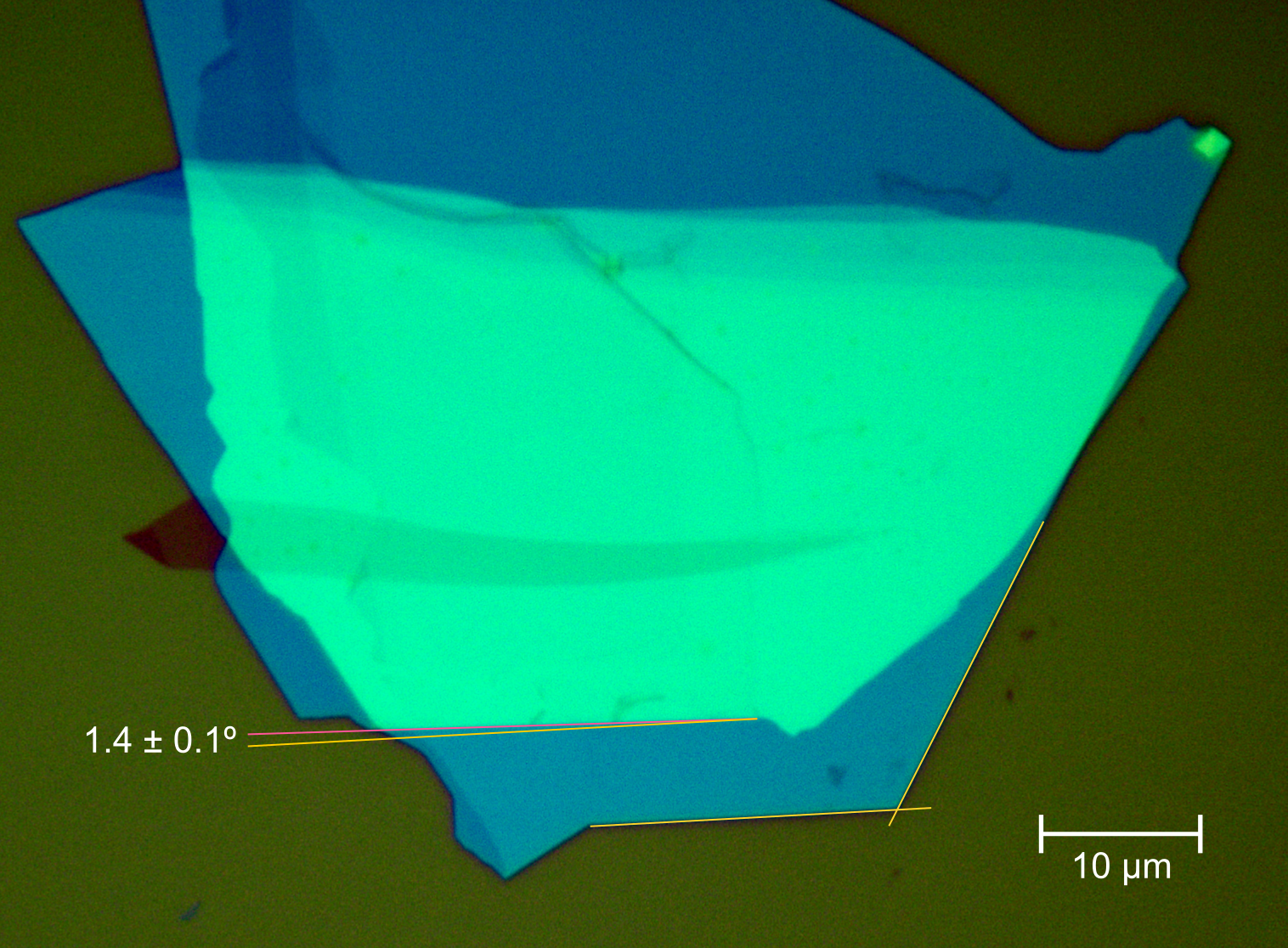}
    \renewcommand{\thefigure}{S1}
    \caption{\textbf{Alignment of BN flakes.} Optical microscope image showing crystallographic axes of top and bottom BN flakes of the stack. The relative angle between long edges is $1.4\pm0.1\degree$.}
    \label{figS1}
\end{figure*}

\end{section}

\newpage

\begin{section}{Additional transport data}

Fig. S2 shows the zoomed-in temperature dependence of four-probe longitudinal resistance $R_\mathrm{xx}$ measurements around the hole (Fig. S2a) and electron (Fig. S2b) superconducting domes. The back gate $V_\mathrm{BG}$ was used to tune the carrier density in one of the two bistable states. Vertical linecuts of $R_\mathrm{xx}$ vs. $T$ for different filling factors $\nu$ are plotted for the hole (Fig. S2c) and electron (Fig. S2d) regions. To extract the critical temperatures $T_\mathrm{c}$, first the normal state resistance $R_\mathrm{N}$ was fit to the equation $R_\mathrm{N}(T) = aT+b$ at higher temperatures. Then, $T_\mathrm{c}$ was extracted from the intersection of the data with $xR_\mathrm{N}$ for $x$ = 0.5, 0.4, and 0.3 for 50\%, 40\%, and 30\% of $R_\mathrm{N}$, respectively. On the hole side, the maximal $T_\mathrm{c}$ occurs at $\nu=-2.62$ with $T_\mathrm{c,50\%}$ = 2.15 K, $T_\mathrm{c,40\%}$ = 1.88 K, and $T_\mathrm{c,30\%}$ = 1.68 K. The electron side has weaker superconductivity where the maximal $T_\mathrm{c}$ occurs at $\nu=2.32$ with $T_\mathrm{c,50\%}$ = 0.83 K, $T_\mathrm{c,40\%}$ = 0.79 K, and $T_\mathrm{c,30\%}$ = 0.74 K.

\vspace{1cm}

\begin{figure*}[!htb]
\includegraphics[width=16 cm]{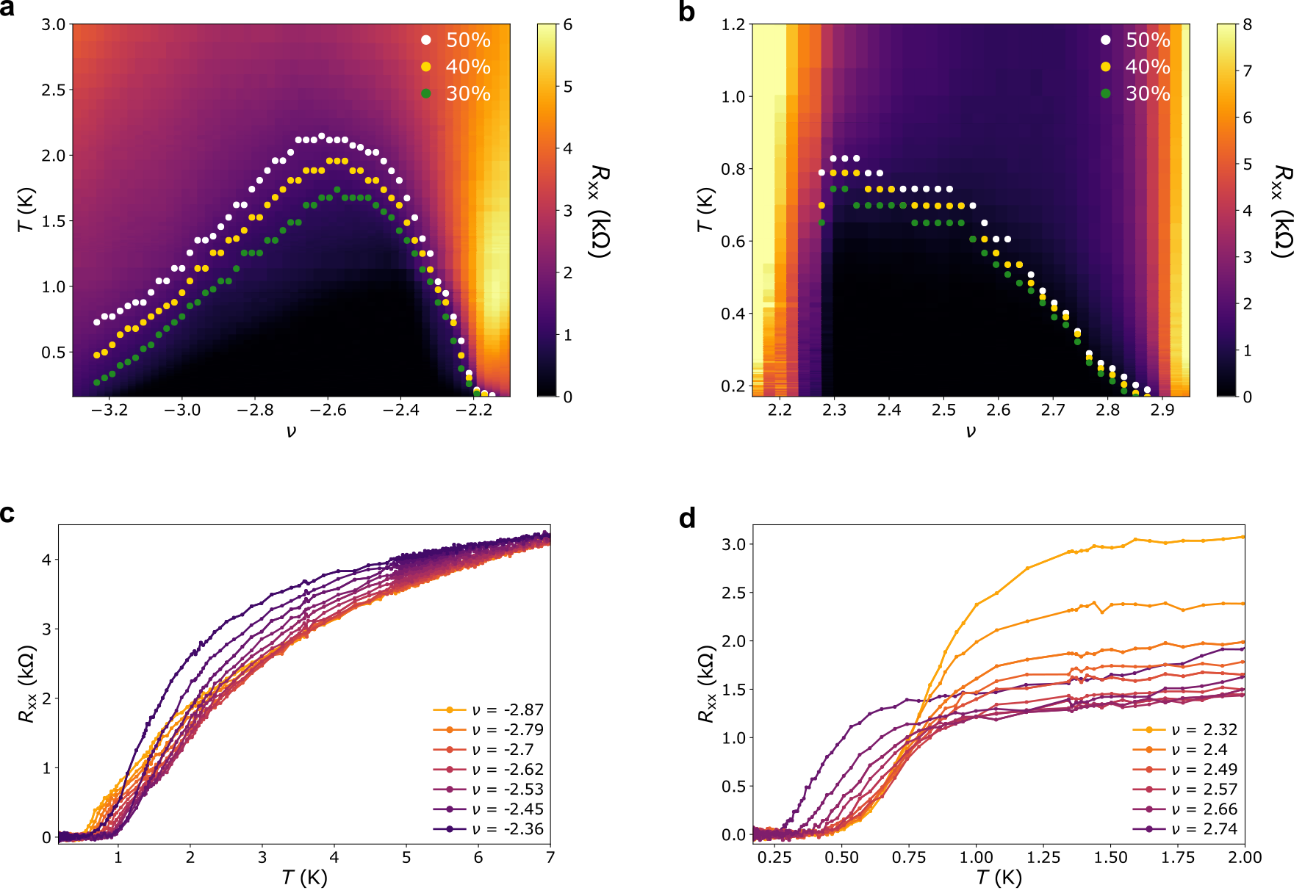}
    \renewcommand{\thefigure}{S2}
    \caption{\textbf{Temperature dependence of superconductivity.}
    \textbf{(a,b)} Four-probe longitudinal resistance $R_\mathrm{xx}$ vs. bottom gate $V_\mathrm{BG}$ and temperature for the hole and electron superconducting domes. The circles denote $T_\mathrm{c}$ extracted from 50\%, 40\%, and 30\% of $R_\mathrm{N}$. \textbf{(c,d)} Vertical linecuts of $R_\mathrm{xx}$ vs. $T$ from \textbf{(a,b)} at different $\nu$.}
    \label{figS2}
\end{figure*}

\begin{figure}[!htb]
\includegraphics[width=10 cm]{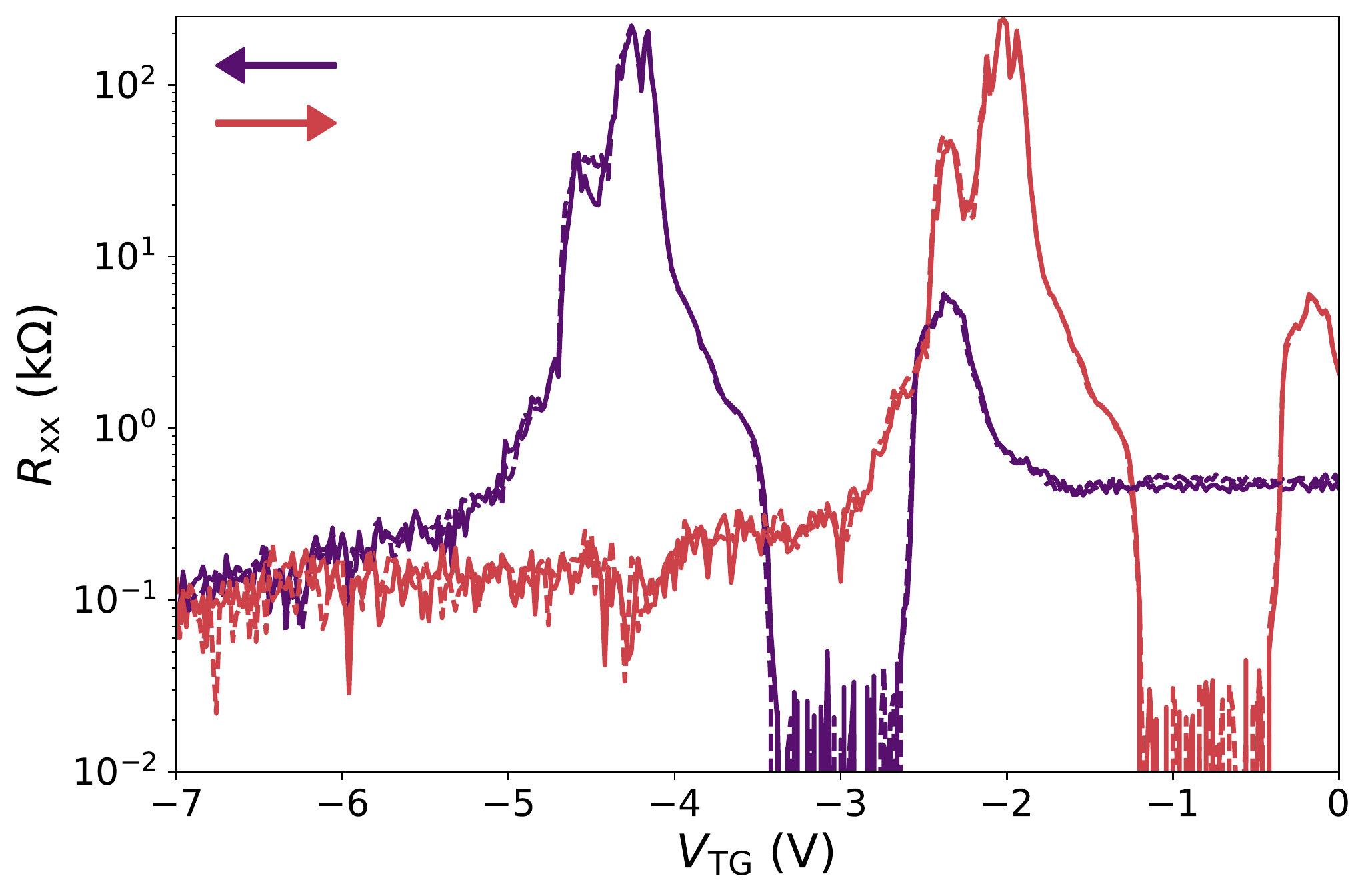}
    \renewcommand{\thefigure}{S3}
    \caption{\textbf{Bistability over time.} Four-probe longitudinal resistance $R_\mathrm{xx}$ versus top gate $V_\mathrm{TG}$. First, the gate was swept down (solid purple) from 0 to $-7$ V and then up (solid pink) back to 0 V. These two scans were then repeated after waiting for 16 hours (dashed lines). The measurements were performed at 65 mK with an AC excitation current of 1 nA.}
    \label{figS3}
\end{figure}

\end{section}

\pagebreak

\begin{section}{Ferroelectric MATBG Device 2}

In addition to the device presented in the main text (Device 1), we fabricated a second MATBG device with aligned top and bottom BN flakes. In this case, the crystallographic edges of the two graphene layers were also closely aligned to the BN flakes (Fig. S4). The top and bottom BN edges are aligned by 0.5$\pm$0.1$\degree$ (modulo 30$\degree$), and the top BN and top monolayer graphene flake in the MATBG stack are aligned by 0.3$\pm$0.1$\degree$ (modulo 30$\degree$). Based on these rotations from optical microscopy and the MATBG twist angle of 0.99$\degree$ extracted from transport measurements (see below), the bottom BN and bottom monolayer graphene flakes are also closely aligned near 1.8$\pm$0.2$\degree$.

\begin{figure}[!htb]
\includegraphics[width=7.5 cm]{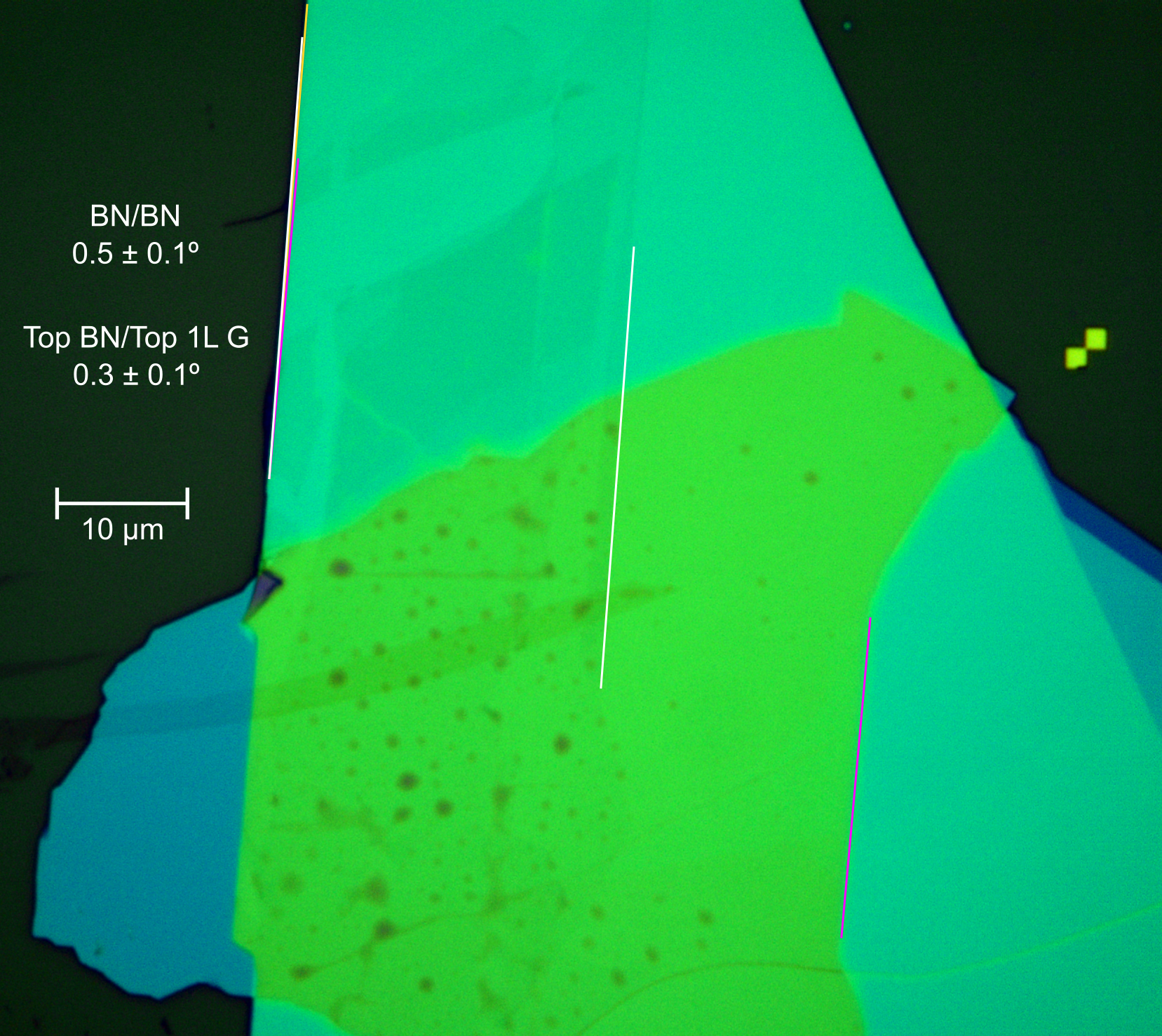}
    \renewcommand{\thefigure}{S4}
    \caption{\textbf{Flake alignment in Device 2.} Optical microscope image showing the angle alignment in a second FE MATBG device. The crystallographic edges of the top (gold) and bottom (purple) BN flakes and the top monolayer graphene (white) are highlighted.}
    \label{figS4}
\end{figure}

Upon sweeping the back gate $V_\mathrm{BG}$ at a temperature of 4.3 K (Fig. S5a), we observe a sharp charge neutrality peak ($\nu$ = 0) and broad correlated insulator peaks at $\nu$ = 2 and 3 in the four-probe longitudinal resistance $R_\mathrm{xx}$. Using the positions of $\nu$ = 0 and 3, we extract an MATBG twist angle of 0.99$\degree$.

Similarly to Device 1, the resistance is strongly dependent on sweep direction of the top gate $V_\mathrm{TG}$ (Fig. S5b). At first, the gate does not appear to inject carriers into MATBG, and then begins to work as a normal field-effect transistor. This hysteresis direction also matches that of Device 1.

\begin{figure}[!htb]
\includegraphics[width=13 cm]{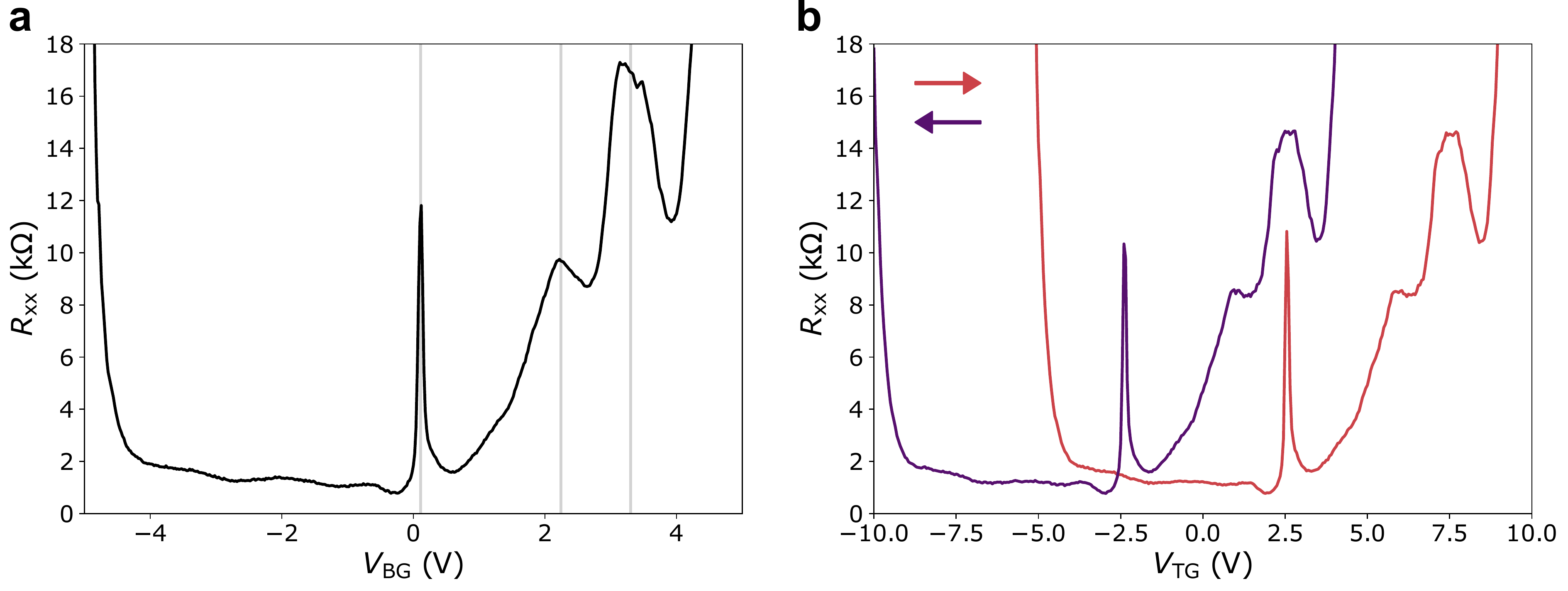}
    \renewcommand{\thefigure}{S5}
    \caption{\textbf{Top and bottom gate sweeps of Device 2.} \textbf{(a)} Four-probe longitudinal resistance $R_\mathrm{xx}$ as a function of bottom gate $V_\mathrm{BG}$. Insulating peaks at $\nu$ = 0, 2, and 3 are indicated by the grey vertical lines. \textbf{(b)} $R_\mathrm{xx}$ as a function of top gate $V_\mathrm{TG}$ when the gate is swept up (pink) and down (purple).}
    \label{figS5}
\end{figure}

Dual-gate $R_\mathrm{xx}$ maps in Figs. S6a and S6b further demonstrate the hysteretic behavior of Device 2. In both cases, $V_\mathrm{BG}$ is swept up as the fast axis. The insulating transport features of $R_\mathrm{xx}$ when the slow axis $V_\mathrm{TG}$ is initially swept up (Fig. S6a) or down (Fig. S6b) are independent of $V_\mathrm{TG}$ until a critical threshold is reached. Beyond this point, the top gate acts normally, characterized by straight-line insulating peaks in the $V_\mathrm{BG}$-$V_\mathrm{TG}$ plane. This behavior can also be visualized in the $n_\mathrm{ext}$-$D_\mathrm{ext}$ plane (Fig. S6c and S6d), as was done for Device 1, using the definitions: $n_\mathrm{ext} = (\varepsilon_\mathrm{BN}/e)(V_\mathrm{TG}/d_\mathrm{TG}+V_\mathrm{BG}/d_\mathrm{BG})$, $D_\mathrm{ext}/\varepsilon_0 = (\varepsilon_\mathrm{BN}/2)(V_\mathrm{TG}/d_\mathrm{TG}-V_\mathrm{BG}/d_\mathrm{BG})$. 

\begin{figure}[!htb]
\includegraphics[width=12 cm]{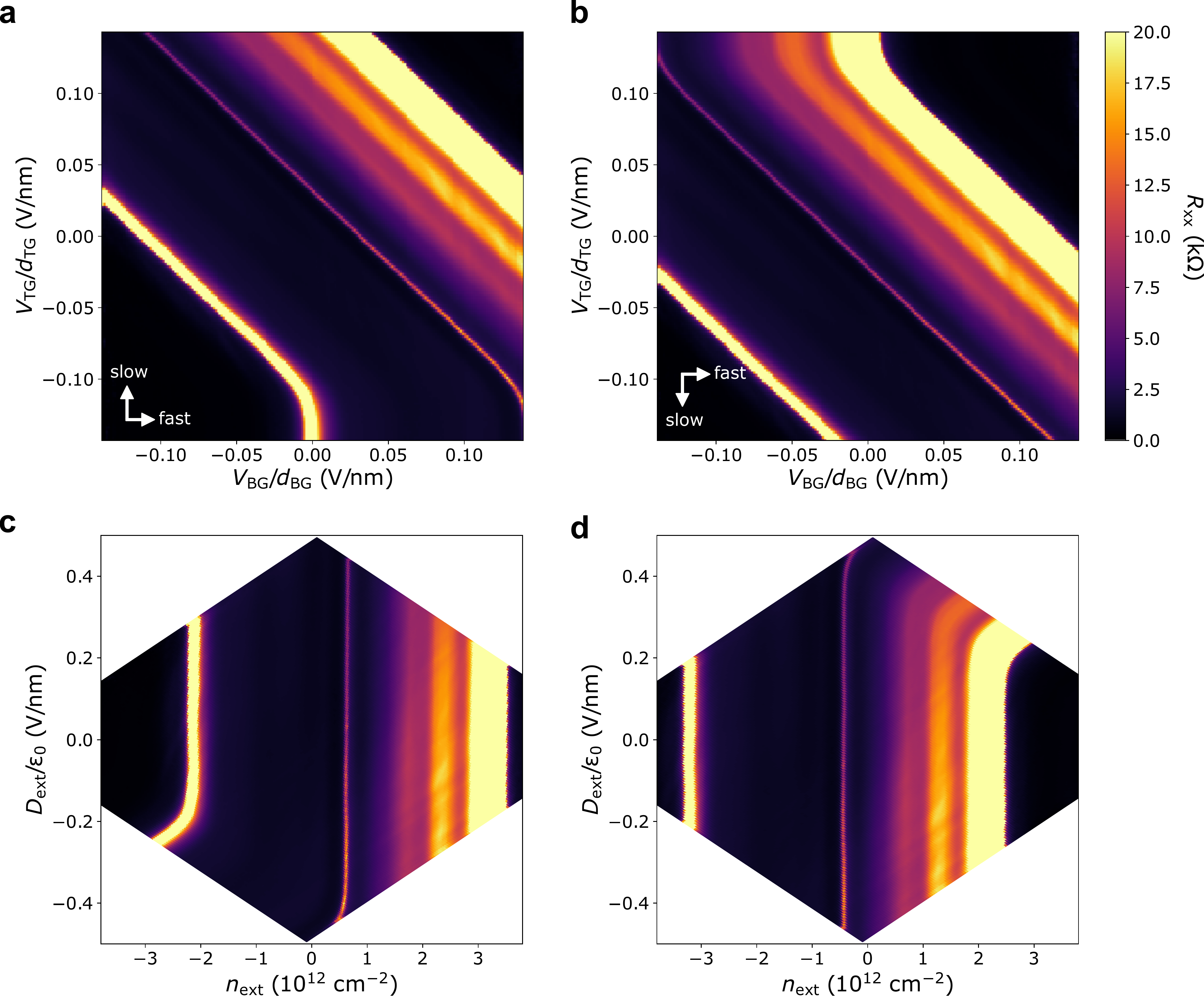}
    \renewcommand{\thefigure}{S6}
    \caption{\textbf{Dual-gate maps of longitudinal resistance.} \textbf{(a, b)} Four-probe longitudinal resistance $R_\mathrm{xx}$ vs. top gate $V_\mathrm{TG}$ and bottom gate $V_\mathrm{BG}$. The fast scan axis is $V_\mathrm{BG}$, swept up from negative to positive. The slow scan axis is $V_\mathrm{TG}$, swept up in \textbf{(a)} and down in $V_\mathrm{TG}$. Axes are normalized to the BN dielectric thicknesses of the two gates. \textbf{(c, d)} Converted maps of \textbf{(a, b)} in units of $n_\mathrm{ext}$ and $D_\mathrm{ext}/\varepsilon_0$ (see text).}
    \label{figS6}
\end{figure}
\newpage
\begin{figure}[!htb]
\includegraphics[width=9 cm]{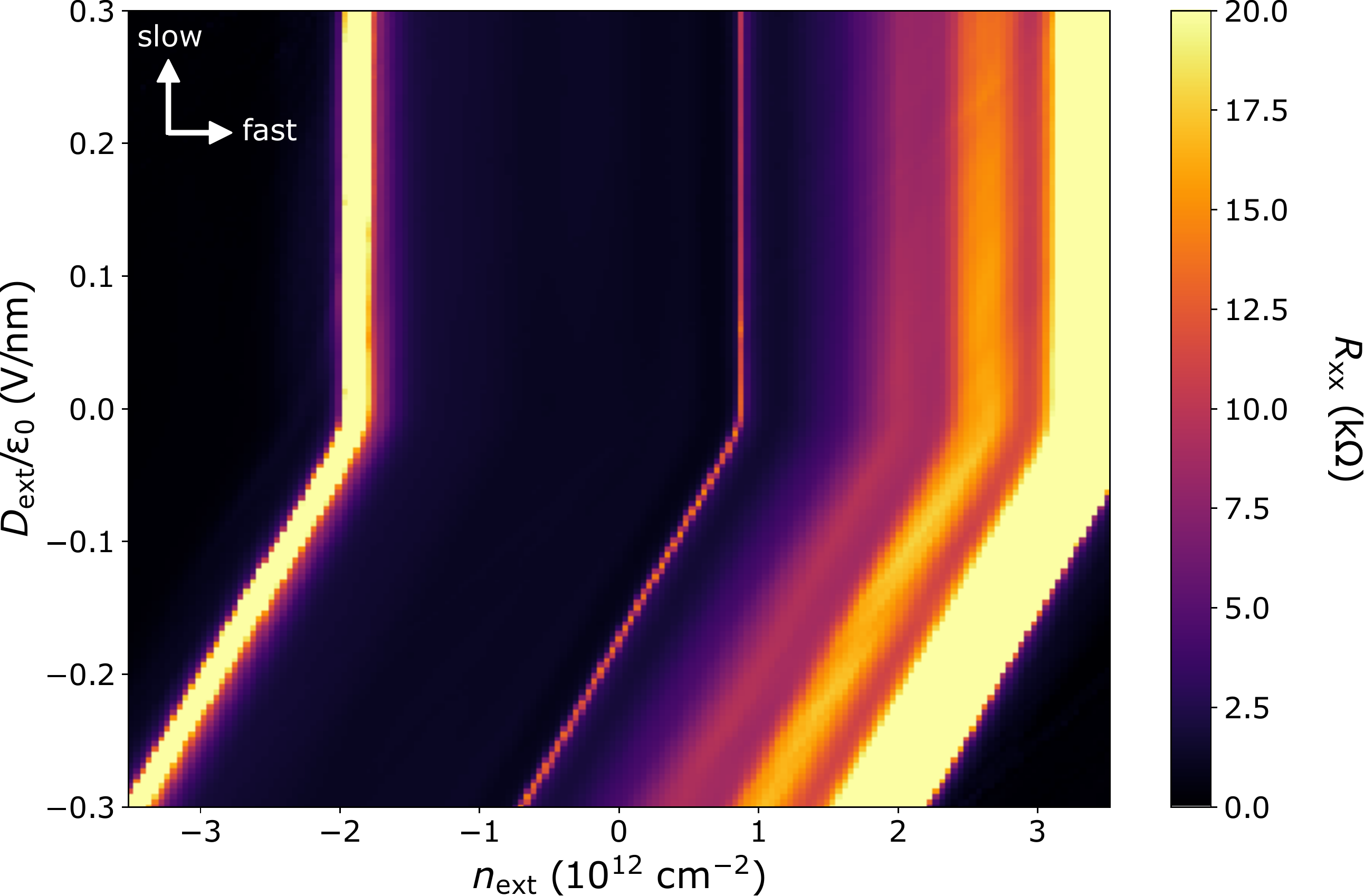}
    \renewcommand{\thefigure}{S7}
    \caption{\textbf{Evolution of $R_\mathrm{xx}$ with applied carrier density and displacement field.} Four-probe longitudinal resistance $R_\mathrm{xx}$ vs. $n_\mathrm{ext}$ and $D_\mathrm{ext}$, as defined in the text from $V_\mathrm{TG}$ and $V_\mathrm{BG}$. The fast scan axis is $n_\mathrm{ext}$ and the slow scan axis is $D_\mathrm{ext}$.}
    \label{figS7}
\end{figure}

In Device 2, we also performed measurements in which we swept $n_\mathrm{ext}$ and $D_\mathrm{ext}$ as the fast and slow axes, respectively (Fig. S7). Initially, at large negative $D_\mathrm{ext}$, the resistive features of MATBG appear to evolve with the defined $D_\mathrm{ext}$. At a certain value of $D_\mathrm{ext}$, however, the system abruptly changes and the correlated states of MATBG are completely independent of $D_\mathrm{ext}$, as expected for an MATBG device without ferroelectric gate behavior.

\end{section}

\bibliography{main.bib}